\documentstyle[12pt]{article}
\title{Exotic smoothness, noncommutative geometry and particle
physics.}
\author{J. S\l adkowski *\\
\sl Institute of Physics, University of Silesia,\\
\sl ul. Uniwersytecka 4, Pl 40-007 Katowice, Poland, \\
and \\
\sl Institute of Physics, University of Wisconsin-Madison,\\
\sl 1150 University Ave., Madison, WI 53706, USA\\}
\begin{document}
\date{}
\baselineskip24pt
\maketitle
\begin{abstract}
\baselineskip24pt
\ \ \ We investigate how exotic differential structures may
reveal themselves in particle physics. The analysis is based on
the A. Connes' construction of the standard model. It is shown that,
if one
of the copies of the spacetime manifold is equipped with an
exotic
differential structure, compact object of geometric origin
may exist
even if the spacetime is topologically trivial. Possible
implications
are discussed. An $SU(3)\otimes SU(2)\otimes U(1)$ gauge model is
constructed. This model may not be realistic but it shows what kind
 of
physical phenomena might be expected due to the existence of exotic
differential structures on the spacetime manifold.
\end{abstract}

\vspace{6mm}

* E-mail: sladk@usctoux1.cto.us.edu.pl
\newpage

\ \ \ There
is no interesting topology on ${\bf R^{4}}$, the Euclidian
four-dimensional
space (or to be more precise it is topologically equivalent to a
single
point space). The counter-intuitive results $^{1-5}$ that ${\bf R^{4}}$
may
be given infinitely many exotic differential structures  raised
question
of their physical consequences $^{6-7}$. An  exotic differential
structure  ${\hat C^{k}}(M)$
on a manifold $M$ is, by definition, a differential structure that is
not diffeomorphic to the one considered as a standard one,
${ C^{k}}(M)$.  This means that the sets od differentiable functions
are
different.  For example, there are  functions on
${\bf R^{4}}$ that are not differentiable on some exotic
${\bf R_{\Theta}^{4}}$ which is homeomorphic but not diffeomorphic
to
${\bf R^{4}}$.
Here we would like to investigate the role that exotic differential
structures on the spacetime manifold may play in particle physics.
Our
starting point will be the A. Connes' noncommutative geometry based
construction of the standard model $^{8-15}$. A. Connes managed to
reformulate the standard notions of differential geometry in a pure
algebraic way that allows to get rid of the differentiability and
continuity requirements. The notion of spacetime manifold $S$ can be
equivalently described by the (commutative) algebra $C^{\infty}(S)$
of smooth
functions on $S$ and can be generalized to (a priori) an arbitrary
noncommutative algebra. Fiber bundles became projective modules
in this
language. A properly generalized connection can describe gauge
fields
on these objects. This allows to incorporate the Higgs field into
the
gauge field so that the correct (that is leading to spontaneously
broken gauge
symmetry) form of the scalar potential is obtained. The reader is
referred to $^{8-15}$ for details.

\ \ \ We shall consider the algebra $A$:

$$ A= M_{1}\left( C^{\infty}\left( S\right) \right) \oplus
M_{2}\left( C^{\infty}\left( S\right) \right) \oplus
M_{1}\left( {\hat C}^{\infty}\left( S\right) \right) \oplus
M_{3}\left( {\hat C}^{\infty}\left( S\right) \right) , \eqno(1)$$
 where  $M_{i}(ring)$ denotes $i\times i$ matrices over the ring
 $C^{\infty}\left( S\right) $  or
${\hat C}^{\infty}\left( S\right) $. The hat denotes that the
functions are smooth with respect to some nonstandard differential
structure on $S$.  The free Dirac operator has the form:

$$D= \pmatrix{
{\not \partial }\otimes Id &
\gamma _{5}\otimes m_{12}&\gamma _{5}\otimes m_{13}&\gamma _{5}
\otimes
m_{14} \cr
\gamma _{5}\otimes m_{21}&{\not \partial }\otimes Id &
\gamma _{5}\otimes m_{23}&\gamma _{5}\otimes m_{24} \cr
\gamma _{5}\otimes m_{31}&\gamma _{5}\otimes m_{32}&
{\hat {\not \partial}} \otimes Id & \gamma _{5}\otimes m_{34} \cr
\gamma _{5}\otimes m_{41}&\gamma _{5}\otimes m_{42}&
\gamma _{5}\otimes m_{43}& {\hat {\not \partial }}\otimes Id \cr},
\eqno(2) $$
here, as before, the hat denotes the "exoticness" of the
appropriate
differential structure. The parameters $m_{ij}$ describe the
fermionic mass
sector. Let $\rho$ be a (self-adjoint) one-form in
$\Omega ^{1}(A) \subset  \Omega ^{*}(A) ,
$ here $\Omega ^{*}(A)$ denotes the universal differential algebra
of $A$ $^{8,9}$:

$$\rho = \sum _{i}  a_{i}db_{i} \ , \ \ a_{i}, b_{i} \varepsilon A.
\eqno(3)$$
We will use the following notation for an $a \varepsilon A$

$$ a = diag \left( a^{1}, a^{2}, a^{3}, a^{4}\right) \eqno(4)$$
with $a^{i}$ belonging to the appropriate matrix algebra in (1). The
physical bosonic fields are defined via the representation $\pi$ in
terms of (bounded) operators in the appropriate Hilbert space $^{6-12}$:

$$\pi \left( a_{0}da_{1} \dots a_{n}\right) = a_{0}
\left[ D,a_{1}\right] \dots \left[ D,a_{n}\right]. \eqno(5)$$
Standard calculations lead to

$$\pi \left( \rho \right) = \pmatrix{
A^{1} &
\gamma _{5} \otimes \phi ^{12} &\gamma _{5} \otimes \phi ^{13}
&\gamma _{5} \otimes \phi ^{14}\cr
\gamma _{5} \otimes \phi ^{21} & A^{2} &
\gamma _{5} \otimes \phi ^{23} &\gamma _{5} \otimes \phi ^{24}\cr
\gamma _{5} \otimes \phi ^{31} &\gamma _{5} \otimes \phi ^{32}&
A^{3} &\gamma _{5} \otimes \phi ^{34}\cr
\gamma _{5} \otimes \phi ^{41} &\gamma _{5} \otimes \phi ^{42}
&\gamma _{5} \otimes \phi ^{43} &A^{4} \cr },
\eqno(6) $$
where

$$A^{p}= \sum _{i} a_{i}^{p}{\not \partial}b_{i}^{p}\ ,\ \ \ \  p=1,2
 \eqno(7a)$$
$$A^{p}= \sum _{i} a_{i}^{p}{{\hat \not \partial }}b_{i}^{p}\ ,
\ \ \ \
p=3,4 \eqno(7b)$$
and

$$\phi ^{pq}= \sum _{i} a_{i}^{p}\left( m_{pq}b_{i}^{q} -
b_{i}^{p}m_{pq}
\right)  \ \ p\neq q .\eqno(8)$$
Note, that the $A^{3}$ and $A^{4}$ are given in terms of the exotic
differential structure. They will be the source of the $SU(3)$
part of
the gauge group. The additional $U(1)$ term $A^{3}$ is the price
we have
to pay for the "exactness" of the $SU(3)$ gauge symmetry:
noncommutative geometry prefers broken gauge symmetries. It is
still an
open question if noncommutative geometry provides us with new
unbroken
symmetries, see Ref. 8-11 for details. There is one subtle step in the
reduction of the gauge symmetry from $SU(2)\otimes U(1)\otimes
U(1)\otimes SU(3)$ to $SU(2)\otimes U(1)\otimes SU(3)$. Namely, one
should require that the $U(1)$ part of the associated connection is
equal to $Y$, the $U(1)$ part of the $SU(3)$ connection and the
"exotic"
$U(1)$ factor is equal to $-Y$.  A more elegant but equivalent
treatment
can be found Ref. 9. But these are defined with respect to different
differential structures! This can be done only locally as the exotic
differential structure defines different set of smooth function
than the standard one (and vice versa).
 We will return to this problem later.  This
defines the algebraic structure of the standard model.  To obtain the
Lagrangian, we have to calculate the curvature $\Theta$, $\Theta =\pi
( d\rho ) = \sum _{i}[D,a_{i}][ D,b_{i}]$. This can be easily done.
The bosonic part of the action is given by the formula

$$ I_{YM}= Tr_{\omega} \left( \Theta ^{2}|D|^{-4} \right) \ ,
\eqno(9)$$
where $Tr_{\omega} $ is the Diximier trace defined by $^{8,9}$

$$Tr_{\omega} \left( |O| \right) = \lim \frac{1}{log N} \sum
_{i=0}^{i=N}\mu _{i}\left( O\right) \ . \eqno(10)$$
Here $\mu _{i}$ denotes the $i-$th eigenvalue of the (compact)
operator
$O$. The Diximier trace gives the logarithmic divergencies, and
gives
zero for operators in the ordinary trace class. We will use the heat
kernel method $^{16-20}$. For a second order positive pseudodifferential
operator $O\ : L^{2}(E)\ ->\ L^{2}(E)$, where $L^{2}(E)$ denotes the
space square integrable functions  on the vector bundle $E$, the
operator

$$ e^{-tO}=\frac{1}{2\pi i}\int _{C} e^{-t\zeta}\left( \zeta Id -O
\right) ^{-1} d\zeta \eqno(11)$$
is well defined for $Re\ t > 0$ $^{16-18}$. Then the Mellin
transformation $^{16}$

$$\int _{0}^{\infty} e^{-tO}t^{s-1}dt=\Gamma \left( s\right) O^{-s}
\eqno(12)$$
provides us with the formula:

$$|D|^{-4}= \int _{0}^{\infty}dt\ te^{-t |D|^2}\ . \eqno(13)$$
Now, we have to restrict ourselves to the case $m_{31}
=m_{32} =m_{41} =m_{42} =m_{13} =m_{14} =m_{23} =m_{24}$ in (2)
so that the free Dirac operator takes the form

$$D= \pmatrix{D_{1} & 0 \cr 0& D_{2}}\ , \eqno(14)$$
where $D_{2}$ is defined with respect to an exotic differential
structure. This allows us to calculate the Diximier trace and
the notion of a point retains its ordinary spacetime sense. This is
not very restrictive as the $SU(3)$ gauge symmetry is unbroken.
Calculation of the Diximier trace in the general case is more
involved (if possible)
and we would loose the convenient spacetime interpretation. The
formula $^{18}$

$$e^{-t\left( D_{1}\oplus D_{2}\right)}= e^{-t\left( D_{1}\right)}
\oplus e^{-t\left( D_{2}\right)} \eqno(15)$$ leads to the following
asymptotic formula:

$$tr\left( \left( f\oplus {\hat f}\right)e^{-t|D|^{2}}\right)=
\int dx^{4}\sqrt{g} f\left( \frac{a_{0}}{t^{2}} +
\frac{a_{1}}{t } +  \dots \right) +
\int {\hat d}x^{4}\sqrt{\hat g} {\hat f}\left( \frac{\hat
a_{0}}{t^{2}} +
\frac{\hat a_{1}}{t } +  \dots \right) , \eqno(16)$$
where $a_{i}$ are the spectral coefficients $^{16-20}$, $g$ is the metric
tensor, dots denote the finite terms
in the limit $t \ -> 0$  and the hat distinguishes between the
standard and exotic structures. For the Dirac Laplace'ans
$|D_{i}|^{2} \ i=1,2$
we have $a_{1}=1$ and $a_{2}$ is equal to the curvature $R$. This
gives the the following value of the Yang-Mills (bosonic) action
(roughly speaking this is the "logarithmic divergence" term):

$$ I_{YM}= \frac{1}{4}
\int dx^{4}\sqrt{g} \ TR\left( \pi ^{2}\left(  \theta
\right) \right) +
\int {\hat d}x^{4}\sqrt{\hat g}\ TR\left( \pi ^{2}\left( {\hat \theta}
\right)  \right)  \ , \eqno(17)$$
where the the trace $TR$ is taken over the Clifford algebra and the
matrix structure. As before, the hat is used to distinguish the
"exotic"
part of the curvature from the "non-exotic" one. Note, that due
to continuity,  the two integrals do not feel the different
differential structures, so formally,
the action looks the same as in the ordinary case. Now, standard
algebraic calculations (after elimination of spurious degrees of
freedom by hand $^{8,10,12-15}$ or by going to the quotient space $^{9}$)
lead to the following Lagrangian (in the Minkowski space):

$$\begin{array}{ll}
{\it L_{YM}}= &\int \sqrt{g}   \lbrace \frac{1}{4} N_{g}\left(
F_{\mu \nu}^{1}F^{1 \mu \nu}  + F_{\mu \nu}^{2}F^{2 \mu \nu} +
F_{\mu \nu}^{c}F^{c \mu \nu} \right) \cr
\ & + \frac{1}{2}Tr\left( mm^{\dag}\right) | \partial
\phi  + A_{1} \phi  - \phi ^{\dag}A_{2} | ^{2}\cr
 \  & -\frac{1}{2} \left( Tr\left( mm^{\dag}\right) ^{2} - \left(
Tr mm^{\dag}\right) ^{2}\right) \left( \phi \phi ^{\dag} -1\right)
^{2}\rbrace d^{4}x\ . \cr
\end{array}\eqno(18)$$
The $SU(3)_{c}$ stress tensor $F_{\mu \nu}^{c}F^{c \mu \nu}$ is
defined with respect to the exotic differential structure.
We will not need the concrete values of the traces in (18) so will
not
quote them (they are analogous to those in $^{21-22}$). Fermion fields
are added in the usual way $^{8-15}$:

$$
\begin{array}{lll}
{\it L_{f}} & = & <\psi | D + \pi \left( \rho \right) | \psi >\cr
\ & = & \int \left( {\bar \psi } _{L} {\not D}\psi _{L} +
{\bar \psi } _{R} {\not D}\psi _{R} + {\bar \psi }_{L}\phi \otimes
m\psi _{R} + {\bar \psi }_{R}\phi ^{\dag} \otimes m ^{\dag}
\psi _{L}\right) d^{4}x \ ,\end{array}\  \eqno(19)$$
where we have included the $\pi (\rho)$ term into ${\not D}$. The
quark
fields are defined with respect to the exotic differential structure.
To proceed, let us review some results concerning exotic differential
structures on ${\bf R ^{4}}$ $^{5-7}$.

\ \ \  An exotic ${\bf R_{\Theta}^{4}}$  consists of a set of points
which can be globally continuously identified with the set four
coordinates $(x^{1}, x^{2},x^{3},x^{4})$. These coordinates may be
smooth
locally but they cannot be globally continued as smooth functions
and no
diffeomorphic image of an exotic ${\bf R_{\Theta}^{4}}$  can be given
such global coordinates in a smooth way. There are uncountable many of
different ${\bf R_{\Theta}^{4}}$. C. H.  Brans has proved the
following theorem $^{7}$:

{\bf Theorem 1.} {\it There exist smooth manifolds which are
homeomorphic but not diffeomorphic to ${\bf R^{4}}$ and for which the
global coordinates $(t,x,y,z)$ are smooth for $x^{2}+y^{2}+z^{2}\geq
a^{2}> 0$, but not globally. Smooth metrics exist for which the
boundary of this region is timelike, so that the exoticness is
spatially
confined.} \\

He has also conjectured that such localized exoticness can act
as an
source for some externally regular field, just as matter or a
wormhole
can. Of course, there are also ${\bf R^{4}_{\Theta}}$  whose
exoticness
cannot be localized. They might have important cosmological
consequences.  We also have $^{7}$

{\bf Theorem 2.} {\it If $M$ is a smooth connected 4-manifolds and
$S$
is a closed submanifold for which $H^{4}(M,S,{\bf Z})=0$, then any
smooth, time-orientable Lorentz metric defined over $S$ can be
smoothly
continued to all of $M$.} \\

\ \ \  Now we are prepared to analyse the Lagrangian given by (18).
Despite the fact that
it looks like an ordinary one we should remember that the
strongly interacting fields are defined with respect to an exotic
differential structure. This means that, in general, these fields
may
not be smooth with respect to the standard differential structure,
although they are smooth solutions with respect to the exotic one.
They
certainly are continuous. In general, only those "exotic" fields that vanish
outside
a compact set (not necessary containing the exotic region) can be expected
to be differentiable with respect to the standard differential structure
(this
is because manifolds are locally Euclidean and constant functions
are differentiable ) and consistent with the
derivation of the Lagrangian (18). Theorem 2. suggests that it might
be
possible to continue a Lorentz structure to all of spacetime so that
(18) make sense (e.g. for a
non-compact manifold $M$, submanifolds $S$ for which $H^{3}(S;{\bf
Z})=0$ fulfil the required conditions $^{7}$).  This means that strongly
interacting fields probably must vanish outside a compact set to be consistent
with the standard (?) differential structure that governs electroweak sector.
One can say
that the exotic geometry confines strongly interacting particles to
live
inside bag-like structures. We do not claim
(although it might be so) that we have found a solution to the
confinement problem, but these results are really astonishing.
Unfortunatelly, the
estimation of the size of such an object is not possible without at the
moment not available information on the global structure of exotic
manifolds. A priori, they may be as small as baryon or as big as
quark star.
What important is is the fact that such object are not black-hole-like
ones. It is
possible to "get inside such an object and go back". There is no
topological obstruction that can prevent us entering the exotic region:
everything is smooth but some fields must have compact supports.
One may investigate its structure as one does in the case of baryons
via electroweak interactions. Of course, the above
analysis is classical: we do not know how to quantize models that
noncommutative geometry provides us with.  Let us conclude by saying
that
exotic differential structures over spacetime may play important role in
particle physics. They may provide us with "confining forces" of pure
geometrical origin: one do not have to introduce additional scalar
fields to obtain bag-like models. We have discussed only exotic
versions of
${\bf R^{4}}$ but there are also other exotic 4-manifolds. The proposed
model is probably far from being a realistic one  but it is the only
one ever
constructed. We have conncted the geometrical exoticness with strong
interactions. We can give only one reason for doing so.
A. Conne's construction provides us with spontaneously broken
gauge symmetries. Exact gauge symmetries
are "out of the way" so we have made the $SU(3)_{color}$ sector
"spatially-exotic".
One can also ask the question {\it if one writes a
Lagrangian, must all of its terms be defined with respect to the same
differential structure on spacetime?} The answer is not so obvious.
Obviously, the topic deserves further investigation. One of the most
important questions is {\it how do exotic differential structures
 influence quantum theory?} This is under investigation. \\

 \ \ \ {\bf Aknowledgement:} I greatly enjoyed the hospitality
 extended
to me during a stay at the Physics Department at the University of
Wisconsin-Madison, where the final version of the paper was discussed
and written down. The stay at Madison was possible due to the
means provided by the {\bf II Joint M. Sk\l odowska-Curie
USA-Poland Fund}.
This work was supported in part by the grant {\bf KBN-PB 2253/2/91}.

\newpage

\section*{References}
\newcounter{bban}
\begin{list}
{\ \arabic{bban} .\ }{\usecounter{bban}\setlength{\rightmargin}
{\leftmargin}}

\item M. Freedman, J. Diff. Geom. {\bf 17}, 357 (1982).
\item S. K. Donaldson, J. Diff. Geom. {\bf 18}, 279 (1983).
\item R. E. Gompf, J. Diff. Geom. {\bf 18}, 317 (1983).
\item S. DeMichelis and M. Freedman, J. Diff. Geom. {\bf 35},
219 (1992).
\item R. E. Gompf, J. Diff. Geom. {\bf 37}, 199 (1993).
\item C. H. Brans and D. Randall, Gen. Rel. Grav. {\bf 25}, 205
(1993).
\item C. H. Brans, IAS-preprint IASSNS-HEP-94/22.
\item A. Connes, Publ. Math. IHES {\bf 62} (1983) 44;{\it
Non-Commutative Geometry} (Academic Press, 1993).
\item J. G. V\'arilly and J. M. Garcia-Bond\'ia, J. Geom. Phys.
{\bf 12}, 223 (1993).
\item A. Connes, in {\it The interface of mathematics and physics}
(Claredon, Oxford, 1990) eds . D. Quillen, G. Segal and S. Tsou.
\item A. Connes and J. Lott, Nucl. Phys. {\bf B} Proc. Suppl.
{\bf 18B}, 29 (1990).
\item A. H. Chamseddine, G. Felder and J. Fr\"ohlich, Phys. Lett.
{\bf B296}, 109 (1992).
\item A. H. Chamseddine, G. Felder and J. Fr\"ohlich, Nucl. Phys.
{\bf B395}, 672 (1993).
\item J. S\l adkowski, to be published in  Int. J. Theor. Phys.,
Bielefeld University preprint, BI-TP93/26 (1993).
\item J. S\l adkowski, Bielefeld University preprint, BI-TP93/64
(1993), Proceedings of the XVII Silesian School
of Theoretical Physics, Szczyrk (1993): Acta Phys. Pol {\bf B} 25 (1994) 1255.
\item P. B. Gilkey, {\it The Index Theorem and the Heat Equation},
(Princeton Univ. Press, Princeton 1984).
\item P. B. Gilkey, {\it Invariance Theory the Heat Equation, and
the Atiyah-Singer Index Theorem}, (Publish or Perish, Wilmington
1984).
\item P. B. Gilkey, Inv. Math. {\bf 26}, 231 (1974).
\item N. Hurt, {\it Geometric Quantization in Action}, (Reidel,
Dordrecht, 1983).
\item R. Ma\' nka and J. S\l adkowski, Phys. Lett. {\bf B224},
97 (1989).
\item D. Kastler and T. Sch\"ucker, Theor. Math. Phys. {\bf 92},
223 (1992); (English translation p. 1075 (1993).
\item J. M. Garcia-Bond\'ia, preprint; hep-th/9404075.

\end{list}

\end{document}